\documentclass[%
 reprint,
nofootinbib,
 amsmath,amssymb,
 aps,
prl,
floatfix
]{revtex4-2}

\usepackage{lmodern}
\usepackage{graphicx}
\usepackage{bm}

\usepackage{xcolor}
\definecolor{blue}{RGB}{0, 0, 128}
\definecolor{darkgray}{RGB}{64, 64, 64}   
\definecolor{teal}{RGB}{0, 100, 100}
\definecolor{darkred}{RGB}{139, 0, 0}

\usepackage{hyperref}
\hypersetup{linktocpage,colorlinks,citecolor={blue},pdfdisplaydoctitle=true,pdfpagemode=UseOutlines,bookmarksnumbered=true,urlcolor={teal},linkcolor={darkred}}
\usepackage{mathrsfs,dsfont}
\usepackage{color}
\usepackage{cleveref}

\newcommand{\bra}[1]{\langle #1 |} 
\newcommand{\ket}[1]{| #1 \rangle } 
\newcommand{\upd}{\mathrm{d}}
\newcommand{\tr}{\mathrm{tr}}
\newcommand{\ie}[0]{\textit{i.e.} }
\newcommand{\eg}[0]{\textit{e.g.} }

\newcommand{\Id}[0]{\mathrm{I}}
\newcommand{\rhof}[0]{\rho_\infty}

\begin{document}

\title{Bootstrapping the stationary state of bosonic open quantum systems}

\author{Gustave Robichon}
\author{Antoine Tilloy}
\email{antoine.tilloy@minesparis.psl.eu}
\affiliation{Laboratoire  de Physique de l’Ecole Normale Supérieure - PSL, Centre Automatique et Systèmes Mines Paris - PSL, CNRS, Inria, PSL Research University, Paris, France}

\begin{abstract}
\noindent
We propose a method to compute expectation values of observables in the stationary state of a (Markovian) bosonic open quantum system. Using a hierarchy of semi-definite relaxations, we obtain finer and finer upper and lower bounds to any expectation value of interest. The bounds are rigorous, robust to stationary state degeneracies, and numerically improve as the occupation number increases on the examples we considered. This makes it adapted to the simulation of stationary states of bosonic qubits and in particular dissipatively stabilized cat qubits.
\end{abstract}

\maketitle

\paragraph*{Introduction --}
We consider quantum systems made of a few bosonic modes interacting with a Markovian bath. Our objective is to describe the stationary state of their dynamics rigorously, as a first step towards a more complete characterization of their real-time and spectral properties. 

Such open quantum systems are the building blocks of some of the most promising approaches to quantum computing where quantum information is encoded in bosonic degrees of freedom \cite{kjaergaard2020_annualreview_superconductingqubits}, that are both degraded and stabilized by their environment. Naturally, there is no hope of characterizing such systems rigorously in the regime where they instantiate a large scale fault-tolerant quantum computer, that is, in the many-body regime and with sufficiently low noise. Nevertheless, it is crucial (and not implausible) to be able to simulate elementary building blocks, made of a few modes, that implement the basic quantum computing routines (simple gates, state preparation, state stabilization, error correction, etc.) we aim to realize.

For a few (say $n\leq 10$) idealized qubits, that are exactly two level systems, classical simulation is easy even in the open case. Indeed, the density matrix $\rho$ of the system is a $2^n \times 2^n$ complex matrix, that easily fits in the memory of a computer. For effective qubits implemented in bosonic systems, direct simulation remains feasible if the infinite dimensional Hilbert space is well approximated by its first $2$ levels, as is \eg the case for transmons. 

Recently, other types of encoding have been proposed that exploit the larger Hilbert space available with bosons to better protect quantum information from typical sources of noise \cite{albert2018_optimizedbosoniccodes,joshi2021_bosonicqubitreview}. One example is the cat qubit \cite{guillaud2023_catlectures}, where the logical $0$ and $1$ states are encoded in the vector space spanned by two coherent states $\ket{+\alpha}$ and $\ket{-\alpha}$. Another example is the Gottesman-Kitaev-Preskill (GKP) qubit \cite{gottesman2001_gkp}, where the logical states are encoded in two grids in phase space. Importantly, improving the intrinsic protection of these qubits mechanically increases the average number of excitations/bosons per mode needed to describe them accurately, and thus makes their classical simulation \emph{a priori} more difficult.

The standard approach is to simply truncate the bosonic Hilbert space up to a maximum number $N_c$ of excitations. In a regime where they offer useful protections, cat qubits may require $N_c\sim 100$ and GKP qubits $N_c \sim 1000$ to get negligible truncation errors. This implies that simulating a system of $3$ effective qubits, with density matrix $N_c^{3}\times N_c^{3}$, is already a difficult task. One possible refinement of this variational approach is to consider a truncation in another basis, or better, to forgo linearity and carve out a well chosen low-dimensional ``compressed'' submanifold of the Hilbert space~\cite{lebris2013_lowrank,schlegel2023_coherentstate,santos2024_lowrank}. It is however not yet clear if this is efficiently doable in all situations of interest.

Further, even if this latter approach eventually tames the computational difficulty of the problem, it does not provide rigorous results, in the form of bounds on the observables of interest. At best, one can confirm that the numerical results seem to converge as the truncation parameter (or submanifold dimension) is increased. For closed systems, the variational approach at least provides a rigorous upper bound on the ground state energy. However, in the open system case, the equivalent quantity is the largest eigenvalue $\lambda_1$ of the Lindblad generator of the dynamics. Generically, with dissipation, $\lambda_1 = 0$ and is thus pointless to upper bound numerically. Nonetheless, we will see that knowing $\lambda_1$ exactly provides a crucial advantage for computations in the open case.

Another orthogonal approach to solve problems in large (or infinite) Hilbert spaces is semi-definite relaxations (SDR) or the \emph{bootstrap}. Instead of an inner approximation of the state $\rho$ on a smaller space, that variational methods provide, one asks directly for an approximation of expectation values $\tr[\mathcal{O}\rho]$ on a larger space than the one \emph{a priori} allowed. Semi-definite relaxations have the advantage that they can be used to obtain rigorous upper and lower bounds on simple observables, and not just a lower bound on the energy. 

SDR have been widely used in quantum information theory \cite{skrzypczyk2023_sdp_qinfo_review}, for example to bound the space of correlations reachable with local measurements in the Navascu\'es-Pironio-Ac\'in hierarchy \cite{navascues2007_NPA_PRL,navascues2008_NPA_NJP}. Relaxations have also been used to constrain quantum field theories of critical phenomena, with the \emph{conformal boostrap} \cite{poland2019_cftboostrapreview} (culminating with the computation of the critical exponents of the Ising model in $3d$ \cite{elshowk2012_boostrapIsingoriginal,rychkov2020_3dising}), and more general massive quantum field theory, with the \emph{S-matrix bootstrap}~\cite{kruczenski2022snowmasswhitepapersmatrix}. This approach was also applied in the simpler context of few \cite{xizhi2020_qmbootstrap,berenstein2023_qmbootstrap} and many-body quantum mechanics \cite{baumgratz2012_sdplowerbound,wang2024certifying}, where it was recently hybridized with tensor network techniques~\cite{kull2024_mpssdp}.

Such relaxations can be used with two different philosophies in mind. One can try to constrain the space of possible theories, by inputting only a minimal amount of constraints imposed by general physical requirements. Or one can input enough constraints to single out a model, and thus use the relaxation to solve it. In the present letter, we are interested in the second option: our objective is not to chart the space of Lindbladians, but to find the stationary state of specific ones.

\paragraph*{The problem --} The Hilbert space $\mathscr{H}$ of our system is that of $n$ bosonic modes, \ie $\mathscr{H} = \bigotimes_{j=1}^n \mathcal{F}^{(j)} $ where $\mathcal{F}^{(j)} =  \text{span}\{\ket{m}_j,~ m\in \mathbb{N}_0\}\cong \mathcal{\ell}^2(\mathbb{N}_0)$. As is customary, we introduce the creation and annihilation operators $a^\dagger_j, a_{j'}$ associated to each factor $\mathcal{F}^{(j)}$, which verify the canonical commutation relations $[a_j,a_{j'}^\dagger] = \delta_{j,j'}$ and $a_j\ket{0}_j = 0$.

We assume that the coupling with the bath is Markovian and thus that the dynamics of the system density matrix $\rho_t$ is given by a master equation of the Gorini-Kossakowski-Sudarshan-Lindblad (GKSL) form \cite{Lindblad1976,gorini1976,breuer2002theory}:
\begin{equation}\label{eq:ME}
\frac{\upd}{\upd t} \rho_t = \mathcal{L}(\rho_t) = -i\,[H,\rho_t] + \sum_{k=1}^r \mathcal{D}[c_k](\rho_t)
\end{equation}
with $\mathcal{L}$ the total Lindbladian, $H = H^\dagger$ the Hamiltonian and $\mathcal{D}[c](\rho) := c \rho c^\dagger - \frac{1}{2} \{c^\dagger c ,\rho\}$ is the \emph{dissipator} associated to $c$. 

In most physical situations of interest, the operators $H$ and $c_k$ appearing in the master equation are low degree polynomials in $a^\dagger_j$ and $a_j$. We focus on the case where the coefficients of these polynomials are time independent.

Given the master equation \eqref{eq:ME}, one is usually interested in i) finding the stationary states $\rho_\infty$, \ie the states such that $\mathcal{L}(\rho_\infty) = 0$, ii) finding the gap $\Delta := -\mathrm{Re}(\lambda_2)$ where $\lambda_2$ is the first non-zero eigenvalue of $\mathcal{L}$, and iii) computing real-time dynamics.
We focus on the first problem here, hoping that it serves as a first step towards solving the other two.

The standard approach is to truncate the Hilbert space, allowing at most $N_c$ excitations above the Fock vacuum. Namely, for each Boson, one considers the truncated Fock space $\mathcal{F}^{(j)}_{N_c}:= \text{span}\{\ket{m}_j, 0 \leq m \leq N_c$\}. All operators are now finite matrices acting on a total Hilbert space of dimension $N_c^n$, the stationary state and gap problems reduce to linear algebra, and real-time evolution to the solution of an ordinary differential equation. Although the truncation error decreases fast, asymptotically, as a function of $N_c$, one needs $N_c$ to be much larger than the typical occupation number \ie $N_c\gg \langle \hat{n}\rangle = \langle a^\dagger a\rangle$ to enter this regime.

\paragraph*{The method --}
We care about the state $\rho_\infty$ only to the extent that it allows us to estimate expectation values $\tr[\mathcal{O} \rho_\infty]$, usually only for a small list of simple (low degree) operators (typically the number operator $n_j = a_j^\dagger a_j$ or a quadrature \eg $x_j=(a_j + a_j^\dagger)/\sqrt{2}$). We may bound expectations of simple operators directly, without trying to expand $\rho_\infty$ itself in a basis.

To be concrete, let $S_D$ be the set of normal-ordered monomials in $a_j,a_j^\dagger$ of total degree at most $D$, \ie $S_D = \{\mathrm{I},a_1,a_1^\dagger,a_2,\dots,a_j^\dagger a_j, \dots \}$. We order the $N_D$ elements in this set in a way compatible with the inclusion $S_{D} \subset S_{D+1}$ and write $\mathcal{O}_i$ the $i$-th element ($\mathcal{O}_1 = \Id$).

Imagine that we managed to write an operator of interest $\mathcal{O}$ \emph{exactly} in the following ``null'' + ``sum of squares'' (SOS) way:
\begin{equation}\label{eq:expansion}
    \mathcal{O} = \alpha \,  \Id + \sum_{i=1}^{N_{D_1}} \beta_i \mathcal{L}^\dagger(\mathcal{O}_i) + \sum_{i,j =1}^{N_{D_2}} \gamma_{ji} \, \mathcal{O}_i^\dagger \mathcal{O}_j
\end{equation}
with $\gamma$ a positive semi-definite (PSD) matrix, $\beta$ a vector, $D_1,D_2$ two (potentially different) maximal degrees and $\mathcal{L}^\dagger$ the adjoint of $\mathcal{L}$, \ie 
\begin{equation}
  \mathcal{L}^\dagger(\mathcal{O}) = i\,[H,\mathcal{O}] + \sum_{k=1}^r \mathcal{D}^\dagger[c_k](\mathcal{O})
\end{equation}
with $\mathcal{D}^\dagger[c](\mathcal{O}) = c^\dagger \mathcal{O} c - \frac{1}{2}\{c^\dagger c,\mathcal{O}\}$.
Then we would have
\begin{align} 
    \tr[\mathcal{O}\rhof] &= \alpha + \sum_{i=1}^{N_{D_1}} \beta_i \underset{\tr[\mathcal{O}_i \mathcal{L}(\rhof)]}{\underbrace{\tr[\mathcal{L}^\dagger(\mathcal{O}_i) \rhof]}}+ \!\sum_{i,j =1}^{N_{D_2}} \! \gamma_{ji}  \underset{M_{ij}~ \text{with}~ M\succeq 0}{\underbrace{\tr[\mathcal{O}_i^\dagger \mathcal{O}_j \rhof]}} \nonumber \\
    &= \alpha + \sum_{i=1}^{N_{D_1}} \beta_i \cdot 0 + \tr[\gamma M]\\
    &\geq \alpha \nonumber
\end{align}
Hence, the equality \eqref{eq:expansion} is a \emph{proof} that $\tr[\mathcal{O}\rho_\infty] \geq \alpha$. Here, it is crucial that $\mathcal{L}(\rhof)$ is exactly zero, which is where the Lindblad setup gets easier than the closed system case where one has to first derive bounds on the smallest eigenvalue $E_0$, $H\ket{\psi_0} = E_0\ket{\psi_0}$, which is not known \cite{wang2024certifying}. Our objective is to construct explicitly such an equality \eqref{eq:expansion}, while maximizing over $\alpha$ to get the best possible bound. Such a problem, as we will see, is a semi-definite program.

Since all polynomials can be put in normal ordered form we can write:
\begin{equation}
 \forall i,j \leq N_{D_2},~~~~  \mathcal{O}_i^\dagger \mathcal{O}_j = \sum_{k=1}^{N_{2D_2}} A_{ij}^k \mathcal{O}_k
\end{equation}
where $A$ contains fixed coefficients that can be computed recursively using the commutation relations.

As we assumed that the Lindbladian is constructed from operators $H$ and $c_k$ that are low degree polynomials in $a_j,a_j^\dagger$ we also have
\begin{equation}
   \forall i \leq N_{D_1},~~~~   \mathcal{L}^\dagger(\mathcal{O}_i) = \sum_{k=1}^{N_{D_1+d_\mathcal{L}}} L^k_i \mathcal{O}_k
\end{equation}
where $d_\mathcal{L}$ is the degree of the superoperator $\mathcal{L}$, \ie the maximum between the degree of $H$ and twice the degree of the $c_k$'s. Finally, the operator $\mathcal{O}$ we care about can also be expanded into operators from the list:
\begin{equation}
    \mathcal{O} = \sum_{k} B_k \mathcal{O}_k
\end{equation}

Finding the largest $\alpha$ consistent with equation \eqref{eq:expansion} thus amounts to solving the following optimization problem:
\begin{equation}\label{eq:sdp_dual}
\begin{split}
   &~~~~~~~~~~ \max_{\alpha,\beta,\gamma} ~~ \alpha  \\
    \text{with}~ &\forall k \geq 0, ~ B^k = \alpha \, \delta^{1k} + \beta_i \, L_i^k + \gamma_{ij} \, A_{ij}^k \\
     &\gamma \succeq 0 ~~.
\end{split}
\end{equation}
Importantly, the constraint in \eqref{eq:sdp_dual} is trivially satisfied for $k \geq \max(N_{D_1 + d_\mathcal{L}}, N_{2 D_2})$, and we only have a finite number of linear equalities to implement. We fix $D_2=D$, $D_1=2D-d_\mathcal{L}$, and call $D$ the maximum degree of the hierarchy. With this fixed $D$, the maximization problem \eqref{eq:sdp_dual} is a semi-definite program (SDP) with a finite number of variables and is efficiently solvable. Note that we do not even need the best solution to get a rigorous bound, we merely require a point satisfying the constraints. 

The SDP \eqref{eq:sdp_dual} is actually the dual (in the sense of convex optimization) of a primal SDP that is closer to the type considered previously for quantum mechanical problems~\cite{xizhi2020_qmbootstrap,berenstein2023_qmbootstrap}:
\begin{equation}
\begin{split}\label{eq:sdp_primal}
&~ \min_{\Lambda,M} ~~ \Lambda_k\, B_k  \\
    \text{with}~& \Lambda_1 = 1 \\
    &\forall i \geq 0, ~ \Lambda_k L_i^k = 0 \\
    &\forall i,j \geq 0,  \Lambda_k A_{ij}^k = M_{ij} \\
    &M \succeq 0 ~~.
    \end{split}
\end{equation}
This primal formulation \eqref{eq:sdp_primal} has a natural interpretation as a \emph{relaxation} of the infinite dimensional minimization  
\begin{equation}
\begin{split}\label{eq:original_problem}
&~ \min_{\rho} ~~ \tr[\mathcal{O}\rho]  \\
    \text{with}~& \tr[\rho] = 1 \\
    &\mathcal{L}(\rho) = 0  \\
    &\rho \succeq 0 ~~,
    \end{split}
\end{equation}
which clearly has the exact value $\tr[\mathcal{O}\rhof]$ as solution. The SDP \eqref{eq:sdp_primal} is a relaxation of the problem \eqref{eq:original_problem} in that we optimize directly the pseudo-expectation values $\Lambda_k \approxeq \tr[\mathcal{O}_k \rhof]$ and $M_{ij}  \approxeq \tr[\mathcal{O}_i^\dagger \mathcal{O}_j \rhof]$ imposing only a finite number of the constraints implied by $\mathcal{L}(\rhof)=0$ and $\rhof\succeq 0$. Hence, because we minimize on a larger convex set than the one allowed, the solution of \eqref{eq:sdp_primal} is a lower bound to the true value. Logically, we could have just started from \eqref{eq:original_problem}, then derived its primal SDP relaxation \eqref{eq:sdp_primal}, then moved on to its dual \eqref{eq:sdp_dual}, and finally rederived what was our SOS starting point \eqref{eq:expansion}.

Solving the SDP \eqref{eq:sdp_dual} or \eqref{eq:sdp_primal} for increasingly larger maximal degrees provides better and better lower bounds to the value of $\tr[\mathcal{O}\rhof]$. We can equivalently obtain upper bounds on the same observable by lower bounding $\tr[-\mathcal{O}\rhof]$.

Semi-definite relaxations applied to the ground state problem of closed bosonic systems can yield analytically trivial bounds (and spuriously non-trivial numerical bounds) \cite{navascues2013_paradox}. One might worry that our approach suffers from similar problems and yields bounds that do not improve as the maximum degree in the SOS representation \eqref{eq:expansion} is increased past a certain point. This is not the case thanks to the $\sum\beta_i \mathcal{L}^\dagger(\mathcal{O}_i)$ term added to the SOS representation. There are two ways to understand this intuitively: i) this term increases the maximum degree that the SOS term can take, which is thus not limited to the degree of the operator we aim to bound ii) Generically, for a non-degenerate stationary state, operators of the form $\mathcal{L}^\dagger(\mathcal{O}_i)$ span the space of operators orthogonal to the identity -- hence as we increase the maximum degree, the equality \eqref{eq:expansion} can be verified with a smaller and smaller SOS term, yielding tighter inequalities.

\begin{figure}
    \centering
    \includegraphics[width=0.99\columnwidth]{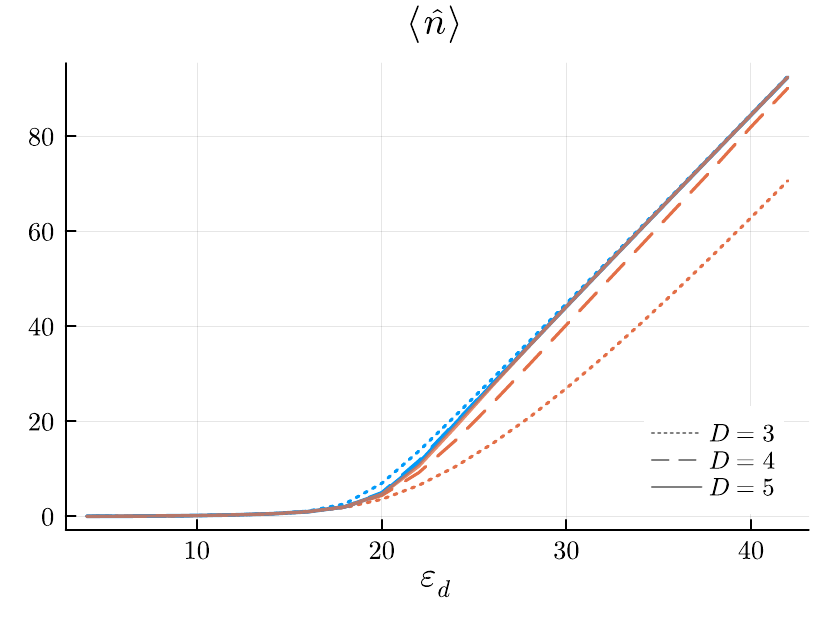}
    \caption{Upper and lower bounds for $ \hat{n} = a^\dagger a$ in the stationary state of the memory-buffer Lindbladian \eqref{eq:memory_buffer} as a function of $\varepsilon_d$, and for maximum degree $D=3,4,5$.}
    \label{fig:multimode}
\end{figure}

\paragraph*{Numerical implementation --}
We specify the problem in \texttt{Julia} using \texttt{JuMP.jl}~\cite{lubin2023_jump}, then solve it with \texttt{Hypatia.jl}~\cite{coey2022hypatia}, which uses a primal-dual interior point method. In our numerical experiments, this solver was the most robust to ill-conditioning of the PSD matrices involved. Depending on the problem, we used double precision (Float64) when sufficient, or arbitrary precision (BigFloats) with as many digits as needed for convergence. Apart from the degree $5$ curves in Fig.~\ref{fig:multimode}, the computations in this letter can be reproduced in a few minutes on a modern laptop.

\paragraph*{Dissipative cat-qubit example --} 
The first example we consider is a single mode undergoing photon loss, two-photon dissipation, and some detuning, described by the Lindbladian:
\begin{equation}\label{eq:lindbladian_stanard_cat}
  \mathcal{L}(\rho) = -i\omega[ a^\dagger a,\rho] + \kappa_1 \,\mathcal{D}[a](\rho) + \kappa_2\, \mathcal{D}\left[a^2 - \alpha^2 \Id\right](\rho)\, ,
\end{equation}
where $\omega$, $\kappa_1$, and $\kappa_2$ are $3$ timescales with $\kappa_2/\kappa_1 \gg 1$ corresponding to the good cat-qubit regime. In this simple model, the stationary state is not degenerate.

\begin{figure}
    \centering
    \includegraphics[width=0.99\columnwidth]{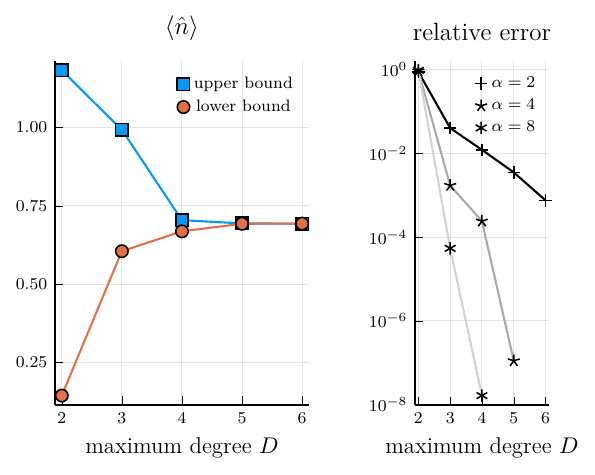}
    \caption{(left) Upper and lower bound on $\langle \hat{n}\rangle$ for the Lindbladian \eqref{eq:lindbladian_stanard_cat}, with $\omega=1, \kappa_1=0.2,\kappa=2, \alpha=1$, as a function of the maximum degree $D$ of the relaxation. (right) Relative error $(\text{upper bound} - \text{lower bound}) /(\text{upper bound} + \text{lower bound})$ as a function of $D$ for increasing values of $\alpha$ and the same $\omega,\kappa_1,\kappa_2$. }
    \label{fig:standard_cat}
\end{figure}

As an illustration, we show the bounds obtained for the operators $\hat{n} = a^\dagger a$, for increasing maximum degrees $D$ and increasing values of $\alpha$ in Fig. \ref{fig:standard_cat}. We observe that the relative error decreases approximately exponentially in the degree, and that larger $\alpha$, hence larger occupation numbers, yield more accurate results.

\begin{figure}
    \centering
    \includegraphics[width=0.975\columnwidth]{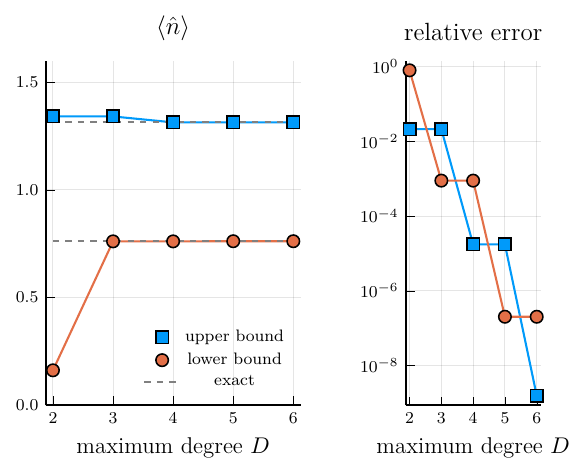}
    \caption{(left) Upper and lower bound on $\langle \hat{n}\rangle$ for the Lindbladian \eqref{eq:lindbladian_stanard_cat}, with $\omega=0, \kappa_1=0,\kappa=2, \alpha=1$, as a function of the maximum degree $D$ of the relaxation. (right) Relative errors on the obtained upper and lower bounds compared to the analytic results, as a function of $D$. }
    \label{fig:perfect_cat}
\end{figure}

\paragraph*{Perfect cat-qubit example --} 
A slightly more subtle example is the singular point $\kappa_1=0, \omega=0$ of the previous example. In that case, the stationary state becomes a stationary subspace of dimension $4$ spanned by superpositions of the coherent states $\ket{\pm\alpha}\bra{\pm\alpha}$. As a result the value of a given observable on the stationary subspace is in general not fixed. Our procedure provides lower bounds and upper bounds that should converge to the extremal values accessible for the observables. This is indeed what we observe numerically in Fig. \ref{fig:perfect_cat}: the stationary state degeneracy yields a gap in the lower and upper bounds as expected.

\paragraph*{A memory-buffer example --} Our approach also works for more complicated systems, with more modes. We consider the so called memory-buffer model, which is a finer description of the cat qubit of the first example, before adiabatic elimination of the buffer. This model was realized experimentally \eg in \cite{berdou2023_100sbitflip} and is given by the Lindbladian
\begin{equation}\label{eq:memory_buffer}
\begin{split}
    \mathcal{L} (\rho) &= - i [H, \rho] + \kappa_a \mathcal{D}[a] \rho + \kappa_b \mathcal{D}[b] (\rho) \\    
    \text{with}~~~ H &= g_2 ( a^2 b^\dagger + a^{\dagger 2} b) + \varepsilon_d( b^\dagger + b)
\end{split}
\end{equation}
where $a$, $b$ are the mode operators of the memory and buffer respectively. We estimate $\langle \hat{n}\rangle = \langle a^\dagger a\rangle$ the population in the memory as a function of $\varepsilon_d$, which is used experimentally to calibrate $g_2$. We choose parameters similar to those of \cite{berdou2023_100sbitflip}: $ g_2=0.25$, $\kappa_b =10\times \kappa_a = 19.4$. An important quantity for this model is $\kappa_2 = 4 g_2^2 / \kappa_b$, which approximately coincides with the $\kappa_2$ of the first example after adiabatic elimination of the buffer. In our case, $\kappa_a / \kappa_2 = 150$, which corresponds to a  qubit limit realized in \cite{berdou2023_100sbitflip}. Results in Fig.~\ref{fig:multimode} show a quick convergence as a function of the maximum degree of the relaxation.

\paragraph*{Discussion --}
We have presented a method to rigorously bound expectation values in the stationary state of bosonic open systems. It is particularly precise at large occupation numbers, in the regime where Hilbert space truncation struggles, and is usable for a few coupled modes. As a side application, our result enables the rigorous contraction of continuous matrix product states (CMPS) \cite{verstraete2010,haegeman2013} with bosonic auxiliary spaces (as considered in \cite{tilloy2019,karanikolaou2021_GaussianCTNS}), \ie CMPS with infinite bond dimension.

We believe there are natural ways to improve precision and go beyond the few mode setup. First, instead of using all normal ordered monomials up to a certain maximal degree, one could use a sparser list better adapted to a given Lindbladian with a simple locality structure (following \eg \cite{wang2024certifying}). Second, we could start from the SOS expansion \eqref{eq:expansion} and parameterize it in a general non-linear way. The resulting optimization problem would no longer be an SDP, and we would lose the related convergence guarantees, but we would still get rigorous bounds.

If high precision is needed, or if very large occupation numbers are considered, ill-conditioning issues make the optimization unstable. This is fairly intuitive, as \eg the primal matrix $M$ of \eqref{eq:sdp_primal} contains expectation values of $a^{\dagger k} a^k$ that are of order $|\alpha|^{2k}$ if the stationary state is close to a coherent state $\ket{\alpha}$. Currently, the only way out we found was to use a solver allowing arbitrary precision, which dramatically increased the computation time and limited us to fairly small SDP. Leveraging some physical intuition about the expected growth of the coefficients, one can likely tame conditioning issues and do better. 

Although the bounds we derived are rigorous, and precise in the examples we considered, a better mathematical understanding of convergence in situations of interest (\eg in the ``good cat qubit" regime) would be useful.

Finally, the most ambitious extension of our approach would be one that goes beyond stationary state estimates. In particular, a method to bound observables for finite time evolution or constrain the spectral gap would be extremely useful. This likely requires other ingredients than the fairly naive ones we have used so far, but is certainly worth exploring.

\begin{acknowledgments}
We thank Ilya Kull, Edoardo Lauria, François Pacaud, and Slava Rychkov for discussions, and are grateful to Pierre Guilmin for suggesting the third example. This work was funded in part by the European Union (ERC, QFT.zip project, Grant Agreement no. 101040260). Views and opinions expressed are however those of the author(s) only and do not necessarily reflect those of the European Union or the European Research Council Executive Agency. Neither the European Union nor the granting authority can be held responsible for them.
\end{acknowledgments}

\bibliography{main}

\begin{thebibliography}{31}%
\makeatletter
\providecommand \@ifxundefined [1]{%
 \@ifx{#1\undefined}
}%
\providecommand \@ifnum [1]{%
 \ifnum #1\expandafter \@firstoftwo
 \else \expandafter \@secondoftwo
 \fi
}%
\providecommand \@ifx [1]{%
 \ifx #1\expandafter \@firstoftwo
 \else \expandafter \@secondoftwo
 \fi
}%
\providecommand \natexlab [1]{#1}%
\providecommand \enquote  [1]{``#1''}%
\providecommand \bibnamefont  [1]{#1}%
\providecommand \bibfnamefont [1]{#1}%
\providecommand \citenamefont [1]{#1}%
\providecommand \href@noop [0]{\@secondoftwo}%
\providecommand \href [0]{\begingroup \@sanitize@url \@href}%
\providecommand \@href[1]{\@@startlink{#1}\@@href}%
\providecommand \@@href[1]{\endgroup#1\@@endlink}%
\providecommand \@sanitize@url [0]{\catcode `\\12\catcode `\$12\catcode `\&12\catcode `\#12\catcode `\^12\catcode `\_12\catcode `\%12\relax}%
\providecommand \@@startlink[1]{}%
\providecommand \@@endlink[0]{}%
\providecommand \url  [0]{\begingroup\@sanitize@url \@url }%
\providecommand \@url [1]{\endgroup\@href {#1}{\urlprefix }}%
\providecommand \urlprefix  [0]{URL }%
\providecommand \Eprint [0]{\href }%
\providecommand \doibase [0]{https://doi.org/}%
\providecommand \selectlanguage [0]{\@gobble}%
\providecommand \bibinfo  [0]{\@secondoftwo}%
\providecommand \bibfield  [0]{\@secondoftwo}%
\providecommand \translation [1]{[#1]}%
\providecommand \BibitemOpen [0]{}%
\providecommand \bibitemStop [0]{}%
\providecommand \bibitemNoStop [0]{.\EOS\space}%
\providecommand \EOS [0]{\spacefactor3000\relax}%
\providecommand \BibitemShut  [1]{\csname bibitem#1\endcsname}%
\let\auto@bib@innerbib\@empty
%</preamble>
\bibitem [{\citenamefont {Kjaergaard}\ \emph {et~al.}(2020)\citenamefont {Kjaergaard}, \citenamefont {Schwartz}, \citenamefont {Braumüller}, \citenamefont {Krantz}, \citenamefont {Wang}, \citenamefont {Gustavsson},\ and\ \citenamefont {Oliver}}]{kjaergaard2020_annualreview_superconductingqubits}%
  \BibitemOpen
  \bibfield  {author} {\bibinfo {author} {\bibfnamefont {M.}~\bibnamefont {Kjaergaard}}, \bibinfo {author} {\bibfnamefont {M.~E.}\ \bibnamefont {Schwartz}}, \bibinfo {author} {\bibfnamefont {J.}~\bibnamefont {Braumüller}}, \bibinfo {author} {\bibfnamefont {P.}~\bibnamefont {Krantz}}, \bibinfo {author} {\bibfnamefont {J.~I.-J.}\ \bibnamefont {Wang}}, \bibinfo {author} {\bibfnamefont {S.}~\bibnamefont {Gustavsson}},\ and\ \bibinfo {author} {\bibfnamefont {W.~D.}\ \bibnamefont {Oliver}},\ }\bibfield  {title} {\bibinfo {title} {Superconducting qubits: Current state of play},\ }\href {https://doi.org/https://doi.org/10.1146/annurev-conmatphys-031119-050605} {\bibfield  {journal} {\bibinfo  {journal} {Annual Review of Condensed Matter Physics}\ }\textbf {\bibinfo {volume} {11}},\ \bibinfo {pages} {369} (\bibinfo {year} {2020})}\BibitemShut {NoStop}%
\bibitem [{\citenamefont {Albert}\ \emph {et~al.}(2018)\citenamefont {Albert}, \citenamefont {Noh}, \citenamefont {Duivenvoorden}, \citenamefont {Young}, \citenamefont {Brierley}, \citenamefont {Reinhold}, \citenamefont {Vuillot}, \citenamefont {Li}, \citenamefont {Shen}, \citenamefont {Girvin}, \citenamefont {Terhal},\ and\ \citenamefont {Jiang}}]{albert2018_optimizedbosoniccodes}%
  \BibitemOpen
  \bibfield  {author} {\bibinfo {author} {\bibfnamefont {V.~V.}\ \bibnamefont {Albert}}, \bibinfo {author} {\bibfnamefont {K.}~\bibnamefont {Noh}}, \bibinfo {author} {\bibfnamefont {K.}~\bibnamefont {Duivenvoorden}}, \bibinfo {author} {\bibfnamefont {D.~J.}\ \bibnamefont {Young}}, \bibinfo {author} {\bibfnamefont {R.~T.}\ \bibnamefont {Brierley}}, \bibinfo {author} {\bibfnamefont {P.}~\bibnamefont {Reinhold}}, \bibinfo {author} {\bibfnamefont {C.}~\bibnamefont {Vuillot}}, \bibinfo {author} {\bibfnamefont {L.}~\bibnamefont {Li}}, \bibinfo {author} {\bibfnamefont {C.}~\bibnamefont {Shen}}, \bibinfo {author} {\bibfnamefont {S.~M.}\ \bibnamefont {Girvin}}, \bibinfo {author} {\bibfnamefont {B.~M.}\ \bibnamefont {Terhal}},\ and\ \bibinfo {author} {\bibfnamefont {L.}~\bibnamefont {Jiang}},\ }\bibfield  {title} {\bibinfo {title} {Performance and structure of single-mode bosonic codes},\ }\href {https://doi.org/10.1103/PhysRevA.97.032346} {\bibfield  {journal} {\bibinfo  {journal} {Phys. Rev. A}\ }\textbf {\bibinfo
  {volume} {97}},\ \bibinfo {pages} {032346} (\bibinfo {year} {2018})}\BibitemShut {NoStop}%
\bibitem [{\citenamefont {Joshi}\ \emph {et~al.}(2021)\citenamefont {Joshi}, \citenamefont {Noh},\ and\ \citenamefont {Gao}}]{joshi2021_bosonicqubitreview}%
  \BibitemOpen
  \bibfield  {author} {\bibinfo {author} {\bibfnamefont {A.}~\bibnamefont {Joshi}}, \bibinfo {author} {\bibfnamefont {K.}~\bibnamefont {Noh}},\ and\ \bibinfo {author} {\bibfnamefont {Y.~Y.}\ \bibnamefont {Gao}},\ }\bibfield  {title} {\bibinfo {title} {Quantum information processing with bosonic qubits in circuit qed},\ }\href {https://doi.org/10.1088/2058-9565/abe989} {\bibfield  {journal} {\bibinfo  {journal} {Quantum Science and Technology}\ }\textbf {\bibinfo {volume} {6}},\ \bibinfo {pages} {033001} (\bibinfo {year} {2021})}\BibitemShut {NoStop}%
\bibitem [{\citenamefont {Guillaud}\ \emph {et~al.}(2023)\citenamefont {Guillaud}, \citenamefont {Cohen},\ and\ \citenamefont {Mirrahimi}}]{guillaud2023_catlectures}%
  \BibitemOpen
  \bibfield  {author} {\bibinfo {author} {\bibfnamefont {J.}~\bibnamefont {Guillaud}}, \bibinfo {author} {\bibfnamefont {J.}~\bibnamefont {Cohen}},\ and\ \bibinfo {author} {\bibfnamefont {M.}~\bibnamefont {Mirrahimi}},\ }\bibfield  {title} {\bibinfo {title} {{Quantum computation with cat qubits}},\ }\href {https://doi.org/10.21468/SciPostPhysLectNotes.72} {\bibfield  {journal} {\bibinfo  {journal} {SciPost Phys. Lect. Notes}\ ,\ \bibinfo {pages} {72}} (\bibinfo {year} {2023})}\BibitemShut {NoStop}%
\bibitem [{\citenamefont {Gottesman}\ \emph {et~al.}(2001)\citenamefont {Gottesman}, \citenamefont {Kitaev},\ and\ \citenamefont {Preskill}}]{gottesman2001_gkp}%
  \BibitemOpen
  \bibfield  {author} {\bibinfo {author} {\bibfnamefont {D.}~\bibnamefont {Gottesman}}, \bibinfo {author} {\bibfnamefont {A.}~\bibnamefont {Kitaev}},\ and\ \bibinfo {author} {\bibfnamefont {J.}~\bibnamefont {Preskill}},\ }\bibfield  {title} {\bibinfo {title} {Encoding a qubit in an oscillator},\ }\href {https://doi.org/10.1103/PhysRevA.64.012310} {\bibfield  {journal} {\bibinfo  {journal} {Phys. Rev. A}\ }\textbf {\bibinfo {volume} {64}},\ \bibinfo {pages} {012310} (\bibinfo {year} {2001})}\BibitemShut {NoStop}%
\bibitem [{\citenamefont {Le~Bris}\ and\ \citenamefont {Rouchon}(2013)}]{lebris2013_lowrank}%
  \BibitemOpen
  \bibfield  {author} {\bibinfo {author} {\bibfnamefont {C.}~\bibnamefont {Le~Bris}}\ and\ \bibinfo {author} {\bibfnamefont {P.}~\bibnamefont {Rouchon}},\ }\bibfield  {title} {\bibinfo {title} {Low-rank numerical approximations for high-dimensional lindblad equations},\ }\href {https://doi.org/10.1103/PhysRevA.87.022125} {\bibfield  {journal} {\bibinfo  {journal} {Phys. Rev. A}\ }\textbf {\bibinfo {volume} {87}},\ \bibinfo {pages} {022125} (\bibinfo {year} {2013})}\BibitemShut {NoStop}%
\bibitem [{\citenamefont {Schlegel}\ \emph {et~al.}(2023)\citenamefont {Schlegel}, \citenamefont {Minganti},\ and\ \citenamefont {Savona}}]{schlegel2023_coherentstate}%
  \BibitemOpen
  \bibfield  {author} {\bibinfo {author} {\bibfnamefont {D.~S.}\ \bibnamefont {Schlegel}}, \bibinfo {author} {\bibfnamefont {F.}~\bibnamefont {Minganti}},\ and\ \bibinfo {author} {\bibfnamefont {V.}~\bibnamefont {Savona}},\ }\href@noop {} {\bibinfo {title} {Coherent-state ladder time-dependent variational principle for open quantum systems}} (\bibinfo {year} {2023}),\ \Eprint {https://arxiv.org/abs/2306.13708} {arXiv:2306.13708 [quant-ph]} \BibitemShut {NoStop}%
\bibitem [{\citenamefont {Santos}\ \emph {et~al.}(2024)\citenamefont {Santos}, \citenamefont {Song},\ and\ \citenamefont {Savona}}]{santos2024_lowrank}%
  \BibitemOpen
  \bibfield  {author} {\bibinfo {author} {\bibfnamefont {S.}~\bibnamefont {Santos}}, \bibinfo {author} {\bibfnamefont {X.}~\bibnamefont {Song}},\ and\ \bibinfo {author} {\bibfnamefont {V.}~\bibnamefont {Savona}},\ }\href@noop {} {\bibinfo {title} {Low-rank variational quantum algorithm for the dynamics of open quantum systems}} (\bibinfo {year} {2024}),\ \Eprint {https://arxiv.org/abs/2403.05908} {arXiv:2403.05908 [quant-ph]} \BibitemShut {NoStop}%
\bibitem [{\citenamefont {Skrzypczyk}\ and\ \citenamefont {Cavalcanti}(2023)}]{skrzypczyk2023_sdp_qinfo_review}%
  \BibitemOpen
  \bibfield  {author} {\bibinfo {author} {\bibfnamefont {P.}~\bibnamefont {Skrzypczyk}}\ and\ \bibinfo {author} {\bibfnamefont {D.}~\bibnamefont {Cavalcanti}},\ }\href {https://doi.org/10.1088/978-0-7503-3343-6} {\emph {\bibinfo {title} {Semidefinite Programming in Quantum Information Science}}},\ 2053-2563\ (\bibinfo  {publisher} {IOP Publishing},\ \bibinfo {year} {2023})\BibitemShut {NoStop}%
\bibitem [{\citenamefont {Navascu\'es}\ \emph {et~al.}(2007)\citenamefont {Navascu\'es}, \citenamefont {Pironio},\ and\ \citenamefont {Ac\'{\i}n}}]{navascues2007_NPA_PRL}%
  \BibitemOpen
  \bibfield  {author} {\bibinfo {author} {\bibfnamefont {M.}~\bibnamefont {Navascu\'es}}, \bibinfo {author} {\bibfnamefont {S.}~\bibnamefont {Pironio}},\ and\ \bibinfo {author} {\bibfnamefont {A.}~\bibnamefont {Ac\'{\i}n}},\ }\bibfield  {title} {\bibinfo {title} {Bounding the set of quantum correlations},\ }\href {https://doi.org/10.1103/PhysRevLett.98.010401} {\bibfield  {journal} {\bibinfo  {journal} {Phys. Rev. Lett.}\ }\textbf {\bibinfo {volume} {98}},\ \bibinfo {pages} {010401} (\bibinfo {year} {2007})}\BibitemShut {NoStop}%
\bibitem [{\citenamefont {Navascués}\ \emph {et~al.}(2008)\citenamefont {Navascués}, \citenamefont {Pironio},\ and\ \citenamefont {Acín}}]{navascues2008_NPA_NJP}%
  \BibitemOpen
  \bibfield  {author} {\bibinfo {author} {\bibfnamefont {M.}~\bibnamefont {Navascués}}, \bibinfo {author} {\bibfnamefont {S.}~\bibnamefont {Pironio}},\ and\ \bibinfo {author} {\bibfnamefont {A.}~\bibnamefont {Acín}},\ }\bibfield  {title} {\bibinfo {title} {A convergent hierarchy of semidefinite programs characterizing the set of quantum correlations},\ }\href {https://doi.org/10.1088/1367-2630/10/7/073013} {\bibfield  {journal} {\bibinfo  {journal} {New Journal of Physics}\ }\textbf {\bibinfo {volume} {10}},\ \bibinfo {pages} {073013} (\bibinfo {year} {2008})}\BibitemShut {NoStop}%
\bibitem [{\citenamefont {Poland}\ \emph {et~al.}(2019)\citenamefont {Poland}, \citenamefont {Rychkov},\ and\ \citenamefont {Vichi}}]{poland2019_cftboostrapreview}%
  \BibitemOpen
  \bibfield  {author} {\bibinfo {author} {\bibfnamefont {D.}~\bibnamefont {Poland}}, \bibinfo {author} {\bibfnamefont {S.}~\bibnamefont {Rychkov}},\ and\ \bibinfo {author} {\bibfnamefont {A.}~\bibnamefont {Vichi}},\ }\bibfield  {title} {\bibinfo {title} {The conformal bootstrap: Theory, numerical techniques, and applications},\ }\href {https://doi.org/10.1103/RevModPhys.91.015002} {\bibfield  {journal} {\bibinfo  {journal} {Rev. Mod. Phys.}\ }\textbf {\bibinfo {volume} {91}},\ \bibinfo {pages} {015002} (\bibinfo {year} {2019})}\BibitemShut {NoStop}%
\bibitem [{\citenamefont {El-Showk}\ \emph {et~al.}(2012)\citenamefont {El-Showk}, \citenamefont {Paulos}, \citenamefont {Poland}, \citenamefont {Rychkov}, \citenamefont {Simmons-Duffin},\ and\ \citenamefont {Vichi}}]{elshowk2012_boostrapIsingoriginal}%
  \BibitemOpen
  \bibfield  {author} {\bibinfo {author} {\bibfnamefont {S.}~\bibnamefont {El-Showk}}, \bibinfo {author} {\bibfnamefont {M.~F.}\ \bibnamefont {Paulos}}, \bibinfo {author} {\bibfnamefont {D.}~\bibnamefont {Poland}}, \bibinfo {author} {\bibfnamefont {S.}~\bibnamefont {Rychkov}}, \bibinfo {author} {\bibfnamefont {D.}~\bibnamefont {Simmons-Duffin}},\ and\ \bibinfo {author} {\bibfnamefont {A.}~\bibnamefont {Vichi}},\ }\bibfield  {title} {\bibinfo {title} {Solving the 3d ising model with the conformal bootstrap},\ }\href {https://doi.org/10.1103/PhysRevD.86.025022} {\bibfield  {journal} {\bibinfo  {journal} {Phys. Rev. D}\ }\textbf {\bibinfo {volume} {86}},\ \bibinfo {pages} {025022} (\bibinfo {year} {2012})}\BibitemShut {NoStop}%
\bibitem [{\citenamefont {Rychkov}(2020)}]{rychkov2020_3dising}%
  \BibitemOpen
  \bibfield  {author} {\bibinfo {author} {\bibfnamefont {S.}~\bibnamefont {Rychkov}},\ }\bibfield  {title} {\bibinfo {title} {3d ising model: a view from the conformal bootstrap island},\ }\href {https://doi.org/10.5802/crphys.23} {\bibfield  {journal} {\bibinfo  {journal} {Comptes Rendus. Physique}\ }\textbf {\bibinfo {volume} {21}},\ \bibinfo {pages} {185} (\bibinfo {year} {2020})}\BibitemShut {NoStop}%
\bibitem [{\citenamefont {Kruczenski}\ \emph {et~al.}(2022)\citenamefont {Kruczenski}, \citenamefont {Penedones},\ and\ \citenamefont {van Rees}}]{kruczenski2022snowmasswhitepapersmatrix}%
  \BibitemOpen
  \bibfield  {author} {\bibinfo {author} {\bibfnamefont {M.}~\bibnamefont {Kruczenski}}, \bibinfo {author} {\bibfnamefont {J.}~\bibnamefont {Penedones}},\ and\ \bibinfo {author} {\bibfnamefont {B.~C.}\ \bibnamefont {van Rees}},\ }\href {https://arxiv.org/abs/2203.02421} {\bibinfo {title} {Snowmass white paper: S-matrix bootstrap}} (\bibinfo {year} {2022}),\ \Eprint {https://arxiv.org/abs/2203.02421} {arXiv:2203.02421 [hep-th]} \BibitemShut {NoStop}%
\bibitem [{\citenamefont {Han}\ \emph {et~al.}(2020)\citenamefont {Han}, \citenamefont {Hartnoll},\ and\ \citenamefont {Kruthoff}}]{xizhi2020_qmbootstrap}%
  \BibitemOpen
  \bibfield  {author} {\bibinfo {author} {\bibfnamefont {X.}~\bibnamefont {Han}}, \bibinfo {author} {\bibfnamefont {S.~A.}\ \bibnamefont {Hartnoll}},\ and\ \bibinfo {author} {\bibfnamefont {J.}~\bibnamefont {Kruthoff}},\ }\bibfield  {title} {\bibinfo {title} {Bootstrapping matrix quantum mechanics},\ }\href {https://doi.org/10.1103/PhysRevLett.125.041601} {\bibfield  {journal} {\bibinfo  {journal} {Phys. Rev. Lett.}\ }\textbf {\bibinfo {volume} {125}},\ \bibinfo {pages} {041601} (\bibinfo {year} {2020})}\BibitemShut {NoStop}%
\bibitem [{\citenamefont {Berenstein}\ and\ \citenamefont {Hulsey}(2023)}]{berenstein2023_qmbootstrap}%
  \BibitemOpen
  \bibfield  {author} {\bibinfo {author} {\bibfnamefont {D.}~\bibnamefont {Berenstein}}\ and\ \bibinfo {author} {\bibfnamefont {G.}~\bibnamefont {Hulsey}},\ }\bibfield  {title} {\bibinfo {title} {Semidefinite programming algorithm for the quantum mechanical bootstrap},\ }\href {https://doi.org/10.1103/PhysRevE.107.L053301} {\bibfield  {journal} {\bibinfo  {journal} {Phys. Rev. E}\ }\textbf {\bibinfo {volume} {107}},\ \bibinfo {pages} {L053301} (\bibinfo {year} {2023})}\BibitemShut {NoStop}%
\bibitem [{\citenamefont {Baumgratz}\ and\ \citenamefont {Plenio}(2012)}]{baumgratz2012_sdplowerbound}%
  \BibitemOpen
  \bibfield  {author} {\bibinfo {author} {\bibfnamefont {T.}~\bibnamefont {Baumgratz}}\ and\ \bibinfo {author} {\bibfnamefont {M.~B.}\ \bibnamefont {Plenio}},\ }\bibfield  {title} {\bibinfo {title} {Lower bounds for ground states of condensed matter systems},\ }\href {https://doi.org/10.1088/1367-2630/14/2/023027} {\bibfield  {journal} {\bibinfo  {journal} {New Journal of Physics}\ }\textbf {\bibinfo {volume} {14}},\ \bibinfo {pages} {023027} (\bibinfo {year} {2012})}\BibitemShut {NoStop}%
\bibitem [{\citenamefont {Wang}\ \emph {et~al.}(2024)\citenamefont {Wang}, \citenamefont {Surace}, \citenamefont {Fr\'{e}rot}, \citenamefont {Legat}, \citenamefont {Renou}, \citenamefont {Magron},\ and\ \citenamefont {Ac\'{i}n}}]{wang2024certifying}%
  \BibitemOpen
  \bibfield  {author} {\bibinfo {author} {\bibfnamefont {J.}~\bibnamefont {Wang}}, \bibinfo {author} {\bibfnamefont {J.}~\bibnamefont {Surace}}, \bibinfo {author} {\bibfnamefont {I.}~\bibnamefont {Fr\'{e}rot}}, \bibinfo {author} {\bibfnamefont {B.}~\bibnamefont {Legat}}, \bibinfo {author} {\bibfnamefont {M.-O.}\ \bibnamefont {Renou}}, \bibinfo {author} {\bibfnamefont {V.}~\bibnamefont {Magron}},\ and\ \bibinfo {author} {\bibfnamefont {A.}~\bibnamefont {Ac\'{i}n}},\ }\href@noop {} {\bibinfo {title} {Certifying ground-state properties of quantum many-body systems}} (\bibinfo {year} {2024}),\ \Eprint {https://arxiv.org/abs/2310.05844} {arXiv:2310.05844 [quant-ph]} \BibitemShut {NoStop}%
\bibitem [{\citenamefont {Kull}\ \emph {et~al.}(2024)\citenamefont {Kull}, \citenamefont {Schuch}, \citenamefont {Dive},\ and\ \citenamefont {Navascu\'es}}]{kull2024_mpssdp}%
  \BibitemOpen
  \bibfield  {author} {\bibinfo {author} {\bibfnamefont {I.}~\bibnamefont {Kull}}, \bibinfo {author} {\bibfnamefont {N.}~\bibnamefont {Schuch}}, \bibinfo {author} {\bibfnamefont {B.}~\bibnamefont {Dive}},\ and\ \bibinfo {author} {\bibfnamefont {M.}~\bibnamefont {Navascu\'es}},\ }\bibfield  {title} {\bibinfo {title} {Lower bounds on ground-state energies of local hamiltonians through the renormalization group},\ }\href {https://doi.org/10.1103/PhysRevX.14.021008} {\bibfield  {journal} {\bibinfo  {journal} {Phys. Rev. X}\ }\textbf {\bibinfo {volume} {14}},\ \bibinfo {pages} {021008} (\bibinfo {year} {2024})}\BibitemShut {NoStop}%
\bibitem [{\citenamefont {Lindblad}(1976)}]{Lindblad1976}%
  \BibitemOpen
  \bibfield  {author} {\bibinfo {author} {\bibfnamefont {G.}~\bibnamefont {Lindblad}},\ }\bibfield  {title} {\bibinfo {title} {On the generators of quantum dynamical semigroups},\ }\href {https://doi.org/10.1007/BF01608499} {\bibfield  {journal} {\bibinfo  {journal} {Communications in Mathematical Physics}\ }\textbf {\bibinfo {volume} {48}},\ \bibinfo {pages} {119} (\bibinfo {year} {1976})}\BibitemShut {NoStop}%
\bibitem [{\citenamefont {Gorini}\ \emph {et~al.}(1976)\citenamefont {Gorini}, \citenamefont {Kossakowski},\ and\ \citenamefont {Sudarshan}}]{gorini1976}%
  \BibitemOpen
  \bibfield  {author} {\bibinfo {author} {\bibfnamefont {V.}~\bibnamefont {Gorini}}, \bibinfo {author} {\bibfnamefont {A.}~\bibnamefont {Kossakowski}},\ and\ \bibinfo {author} {\bibfnamefont {E.~C.~G.}\ \bibnamefont {Sudarshan}},\ }\bibfield  {title} {\bibinfo {title} {{Completely positive dynamical semigroups of N‐level systems}},\ }\href {https://doi.org/10.1063/1.522979} {\bibfield  {journal} {\bibinfo  {journal} {Journal of Mathematical Physics}\ }\textbf {\bibinfo {volume} {17}},\ \bibinfo {pages} {821} (\bibinfo {year} {1976})},\ \Eprint {https://arxiv.org/abs/https://pubs.aip.org/aip/jmp/article-pdf/17/5/821/19090720/821\_1\_online.pdf} {https://pubs.aip.org/aip/jmp/article-pdf/17/5/821/19090720/821\_1\_online.pdf} \BibitemShut {NoStop}%
\bibitem [{\citenamefont {Breuer}\ and\ \citenamefont {Petruccione}(2002)}]{breuer2002theory}%
  \BibitemOpen
  \bibfield  {author} {\bibinfo {author} {\bibfnamefont {H.-P.}\ \bibnamefont {Breuer}}\ and\ \bibinfo {author} {\bibfnamefont {F.}~\bibnamefont {Petruccione}},\ }\href@noop {} {\emph {\bibinfo {title} {The theory of open quantum systems}}}\ (\bibinfo  {publisher} {Oxford University Press, USA},\ \bibinfo {year} {2002})\BibitemShut {NoStop}%
\bibitem [{\citenamefont {Navascués}\ \emph {et~al.}(2013)\citenamefont {Navascués}, \citenamefont {García-Sáez}, \citenamefont {Acín}, \citenamefont {Pironio},\ and\ \citenamefont {Plenio}}]{navascues2013_paradox}%
  \BibitemOpen
  \bibfield  {author} {\bibinfo {author} {\bibfnamefont {M.}~\bibnamefont {Navascués}}, \bibinfo {author} {\bibfnamefont {A.}~\bibnamefont {García-Sáez}}, \bibinfo {author} {\bibfnamefont {A.}~\bibnamefont {Acín}}, \bibinfo {author} {\bibfnamefont {S.}~\bibnamefont {Pironio}},\ and\ \bibinfo {author} {\bibfnamefont {M.~B.}\ \bibnamefont {Plenio}},\ }\bibfield  {title} {\bibinfo {title} {A paradox in bosonic energy computations via semidefinite programming relaxations},\ }\href {https://doi.org/10.1088/1367-2630/15/2/023026} {\bibfield  {journal} {\bibinfo  {journal} {New Journal of Physics}\ }\textbf {\bibinfo {volume} {15}},\ \bibinfo {pages} {023026} (\bibinfo {year} {2013})}\BibitemShut {NoStop}%
\bibitem [{\citenamefont {Lubin}\ \emph {et~al.}(2023)\citenamefont {Lubin}, \citenamefont {Dowson}, \citenamefont {{Dias Garcia}}, \citenamefont {Huchette}, \citenamefont {Legat},\ and\ \citenamefont {Vielma}}]{lubin2023_jump}%
  \BibitemOpen
  \bibfield  {author} {\bibinfo {author} {\bibfnamefont {M.}~\bibnamefont {Lubin}}, \bibinfo {author} {\bibfnamefont {O.}~\bibnamefont {Dowson}}, \bibinfo {author} {\bibfnamefont {J.}~\bibnamefont {{Dias Garcia}}}, \bibinfo {author} {\bibfnamefont {J.}~\bibnamefont {Huchette}}, \bibinfo {author} {\bibfnamefont {B.}~\bibnamefont {Legat}},\ and\ \bibinfo {author} {\bibfnamefont {J.~P.}\ \bibnamefont {Vielma}},\ }\bibfield  {title} {\bibinfo {title} {{JuMP} 1.0: {R}ecent improvements to a modeling language for mathematical optimization},\ }\href {https://doi.org/10.1007/s12532-023-00239-3} {\bibfield  {journal} {\bibinfo  {journal} {Mathematical Programming Computation}\ }\textbf {\bibinfo {volume} {15}},\ \bibinfo {pages} {581–589} (\bibinfo {year} {2023})}\BibitemShut {NoStop}%
\bibitem [{\citenamefont {Coey}\ \emph {et~al.}(2022)\citenamefont {Coey}, \citenamefont {Kapelevich},\ and\ \citenamefont {Vielma}}]{coey2022hypatia}%
  \BibitemOpen
  \bibfield  {author} {\bibinfo {author} {\bibfnamefont {C.}~\bibnamefont {Coey}}, \bibinfo {author} {\bibfnamefont {L.}~\bibnamefont {Kapelevich}},\ and\ \bibinfo {author} {\bibfnamefont {J.~P.}\ \bibnamefont {Vielma}},\ }\bibfield  {title} {\bibinfo {title} {Solving natural conic formulations with {H}ypatia.jl},\ }\href {https://doi.org/https://doi.org/10.1287/ijoc.2022.1202} {\bibfield  {journal} {\bibinfo  {journal} {INFORMS Journal on Computing}\ }\textbf {\bibinfo {volume} {34}},\ \bibinfo {pages} {2686} (\bibinfo {year} {2022})}\BibitemShut {NoStop}%
\bibitem [{\citenamefont {Berdou}\ \emph {et~al.}(2023)\citenamefont {Berdou}, \citenamefont {Murani}, \citenamefont {R\'eglade}, \citenamefont {Smith}, \citenamefont {Villiers}, \citenamefont {Palomo}, \citenamefont {Rosticher}, \citenamefont {Denis}, \citenamefont {Morfin}, \citenamefont {Delbecq}, \citenamefont {Kontos}, \citenamefont {Pankratova}, \citenamefont {Rautschke}, \citenamefont {Peronnin}, \citenamefont {Sellem}, \citenamefont {Rouchon}, \citenamefont {Sarlette}, \citenamefont {Mirrahimi}, \citenamefont {Campagne-Ibarcq}, \citenamefont {Jezouin}, \citenamefont {Lescanne},\ and\ \citenamefont {Leghtas}}]{berdou2023_100sbitflip}%
  \BibitemOpen
  \bibfield  {author} {\bibinfo {author} {\bibfnamefont {C.}~\bibnamefont {Berdou}}, \bibinfo {author} {\bibfnamefont {A.}~\bibnamefont {Murani}}, \bibinfo {author} {\bibfnamefont {U.}~\bibnamefont {R\'eglade}}, \bibinfo {author} {\bibfnamefont {W.}~\bibnamefont {Smith}}, \bibinfo {author} {\bibfnamefont {M.}~\bibnamefont {Villiers}}, \bibinfo {author} {\bibfnamefont {J.}~\bibnamefont {Palomo}}, \bibinfo {author} {\bibfnamefont {M.}~\bibnamefont {Rosticher}}, \bibinfo {author} {\bibfnamefont {A.}~\bibnamefont {Denis}}, \bibinfo {author} {\bibfnamefont {P.}~\bibnamefont {Morfin}}, \bibinfo {author} {\bibfnamefont {M.}~\bibnamefont {Delbecq}}, \bibinfo {author} {\bibfnamefont {T.}~\bibnamefont {Kontos}}, \bibinfo {author} {\bibfnamefont {N.}~\bibnamefont {Pankratova}}, \bibinfo {author} {\bibfnamefont {F.}~\bibnamefont {Rautschke}}, \bibinfo {author} {\bibfnamefont {T.}~\bibnamefont {Peronnin}}, \bibinfo {author} {\bibfnamefont {L.-A.}\ \bibnamefont {Sellem}}, \bibinfo {author} {\bibfnamefont {P.}~\bibnamefont
  {Rouchon}}, \bibinfo {author} {\bibfnamefont {A.}~\bibnamefont {Sarlette}}, \bibinfo {author} {\bibfnamefont {M.}~\bibnamefont {Mirrahimi}}, \bibinfo {author} {\bibfnamefont {P.}~\bibnamefont {Campagne-Ibarcq}}, \bibinfo {author} {\bibfnamefont {S.}~\bibnamefont {Jezouin}}, \bibinfo {author} {\bibfnamefont {R.}~\bibnamefont {Lescanne}},\ and\ \bibinfo {author} {\bibfnamefont {Z.}~\bibnamefont {Leghtas}},\ }\bibfield  {title} {\bibinfo {title} {One hundred second bit-flip time in a two-photon dissipative oscillator},\ }\href {https://doi.org/10.1103/PRXQuantum.4.020350} {\bibfield  {journal} {\bibinfo  {journal} {PRX Quantum}\ }\textbf {\bibinfo {volume} {4}},\ \bibinfo {pages} {020350} (\bibinfo {year} {2023})}\BibitemShut {NoStop}%
\bibitem [{\citenamefont {Verstraete}\ and\ \citenamefont {Cirac}(2010)}]{verstraete2010}%
  \BibitemOpen
  \bibfield  {author} {\bibinfo {author} {\bibfnamefont {F.}~\bibnamefont {Verstraete}}\ and\ \bibinfo {author} {\bibfnamefont {J.~I.}\ \bibnamefont {Cirac}},\ }\bibfield  {title} {\bibinfo {title} {Continuous matrix product states for quantum fields},\ }\href {https://doi.org/10.1103/PhysRevLett.104.190405} {\bibfield  {journal} {\bibinfo  {journal} {Phys. Rev. Lett.}\ }\textbf {\bibinfo {volume} {104}},\ \bibinfo {pages} {190405} (\bibinfo {year} {2010})}\BibitemShut {NoStop}%
\bibitem [{\citenamefont {Haegeman}\ \emph {et~al.}(2013)\citenamefont {Haegeman}, \citenamefont {Cirac}, \citenamefont {Osborne},\ and\ \citenamefont {Verstraete}}]{haegeman2013}%
  \BibitemOpen
  \bibfield  {author} {\bibinfo {author} {\bibfnamefont {J.}~\bibnamefont {Haegeman}}, \bibinfo {author} {\bibfnamefont {J.~I.}\ \bibnamefont {Cirac}}, \bibinfo {author} {\bibfnamefont {T.~J.}\ \bibnamefont {Osborne}},\ and\ \bibinfo {author} {\bibfnamefont {F.}~\bibnamefont {Verstraete}},\ }\bibfield  {title} {\bibinfo {title} {Calculus of continuous matrix product states},\ }\href {https://doi.org/10.1103/PhysRevB.88.085118} {\bibfield  {journal} {\bibinfo  {journal} {Phys. Rev. B}\ }\textbf {\bibinfo {volume} {88}},\ \bibinfo {pages} {085118} (\bibinfo {year} {2013})}\BibitemShut {NoStop}%
\bibitem [{\citenamefont {Tilloy}\ and\ \citenamefont {Cirac}(2019)}]{tilloy2019}%
  \BibitemOpen
  \bibfield  {author} {\bibinfo {author} {\bibfnamefont {A.}~\bibnamefont {Tilloy}}\ and\ \bibinfo {author} {\bibfnamefont {J.~I.}\ \bibnamefont {Cirac}},\ }\bibfield  {title} {\bibinfo {title} {Continuous tensor network states for quantum fields},\ }\href {https://doi.org/10.1103/PhysRevX.9.021040} {\bibfield  {journal} {\bibinfo  {journal} {Phys. Rev. X}\ }\textbf {\bibinfo {volume} {9}},\ \bibinfo {pages} {021040} (\bibinfo {year} {2019})}\BibitemShut {NoStop}%
\bibitem [{\citenamefont {Karanikolaou}\ \emph {et~al.}(2021)\citenamefont {Karanikolaou}, \citenamefont {Emonts},\ and\ \citenamefont {Tilloy}}]{karanikolaou2021_GaussianCTNS}%
  \BibitemOpen
  \bibfield  {author} {\bibinfo {author} {\bibfnamefont {T.~D.}\ \bibnamefont {Karanikolaou}}, \bibinfo {author} {\bibfnamefont {P.}~\bibnamefont {Emonts}},\ and\ \bibinfo {author} {\bibfnamefont {A.}~\bibnamefont {Tilloy}},\ }\bibfield  {title} {\bibinfo {title} {Gaussian continuous tensor network states for simple bosonic field theories},\ }\href {https://doi.org/10.1103/PhysRevResearch.3.023059} {\bibfield  {journal} {\bibinfo  {journal} {Phys. Rev. Res.}\ }\textbf {\bibinfo {volume} {3}},\ \bibinfo {pages} {023059} (\bibinfo {year} {2021})}\BibitemShut {NoStop}%
\end{thebibliography}%

\end{document}